# Theory of Correlated Pairs of Electrons Oscillating in Resonant Quantum States to Reach the Critical Temperature in a Metal

R. Riera and R. A. Rosas-Cabrera

Departamento de Investigación en Física, Universidad de Sonora

Apartado Postal 5-088, 83190 Hermosillo, Sonora, México

R. Rosas

Departamento de Física, Universidad de Sonora

Apartado Postal 1626, 83000 Hermosillo, Sonora, México

Re. Betancourt-Riera and Ri. Betancourt-Riera

Instituto Tecnológico de Hermosillo, Avenida Tecnológico S/N,

Colonia Sahuaro, C.P. 83170, Hermosillo, Sonora, México

## I. ABSTRACT

The formation of Correlated Electron Pairs Oscillating around the Fermi level in Resonant Quantum States (CEPO-RQS), when a metal is cooled to its critical temperature $T = T_C$, is studied. The necessary conditions for the existence of CEPO-RQS are analyzed. The participation of electron-electron interaction screened by an electron dielectric constant of the form proposed by Thomas Fermi is considered and a physical meaning for the electron-phonon-electron interaction in the formation of the CEPO-RQS is given. The internal state of the CEPO-RQS is characterized by taking into account the internal Hamiltonian, obtaining a general equation that represents its binding energy and depends only on temperature and critical temperature. A parameter is also defined that contains the properties that qualitatively characterizes the nature of a material to form the CEPO-RQS.

## II. INTRODUCTION

In the literature the formation of electron pairs in a conductive material due to electron-phonon interaction and coulombian electron-electron repulsion is very important in explaining the theory of superconductivity [1]. The attraction between electrons is interpreted and explained by Cooper [2], considering the screening of the electron-electron interaction by a total dielectric constant due to phonons and electrons. In [2] the condition that the energy of the phonon must be greater than the energy difference between the pair of electrons was imposed in such a way that the screened coulombian repulsion is negative, interpreting this sign as an attraction between the two electrons; these pairs are the carriers of superconductivity in the BCS theory and are responsible for the gap in the energy spectrum [1-6]. Other works [4, 6, 8, 9] attempted to correct the equation that gives the critical temperature in the BCS using many body theories, however, there are no works that calculate other physical magnitudes characteristic of the superconductors based on the expression of the gap and the critical temperature as the BCS does.

In this work we establish the necessary conditions to form Oscillating Correlated Electron Pairs in what we have called Resonant Quantum States (CEPO-RQS) when a metal is cooled to the critical temperature. We have considered the coulombian repulsion interaction screened by an electron dielectric constant of the form proposed by Thomas Fermi, and the electron-phonon-electron interaction described in the Hamiltonian by an oscillatory potential that allows electrons to be trapped in resonant quantum states above and below the Fermi level in the conduction band.

When the critical temperature is reached, the energy difference between the electrons is the same as the energy of the phonon and that is the condition for the pair of electrons, one below and one above the Fermi level with opposite momenta differing by the phonon momentum, oscillate. Thus a change of sign of the momentum of the phonon occurs in each half oscillation.

In this paper we present novel and original results for the case when a metal is cooled to the critical temperature, the necessary conditions are established for the formation of CEPO-RQS and it is shown that the Hamiltonian of the interaction doesn't have self-states, but its square it does.

### III. PHYSICAL MODEL

In Figure 1 we present a conduction band diagram of a metal which reaches its critical temperature $T = T_C$; in the vertical axis we have the energy $E(k)$ and in the horizontal the momentum $k$. We consider the initial state of the pair of electrons in positions 1 and 2 with momenta $k+q$ y $-k$, respectively, and two holes in positions 1' and 2' with momenta $k$ y $-k-q$. The energy difference between states 1 and 2, and 2' and 1' is the energy of a phonon, $\hbar\omega_q$, which is a function of the temperature for $0 \leq T \leq T_C$. These states are located one above and one below the Fermi level ($\varepsilon_f$). Before the temperature reaches its critical value $T = T_C$ the Hamiltonian of the metal is

$$H = -\frac{\hbar^2 \nabla^2}{2m_e} + V_C \tag{1}$$

where the first term is the kinetic energy operator and the second one is the potential energy of the crystal, which for electrons in the conduction band is relatively weak.

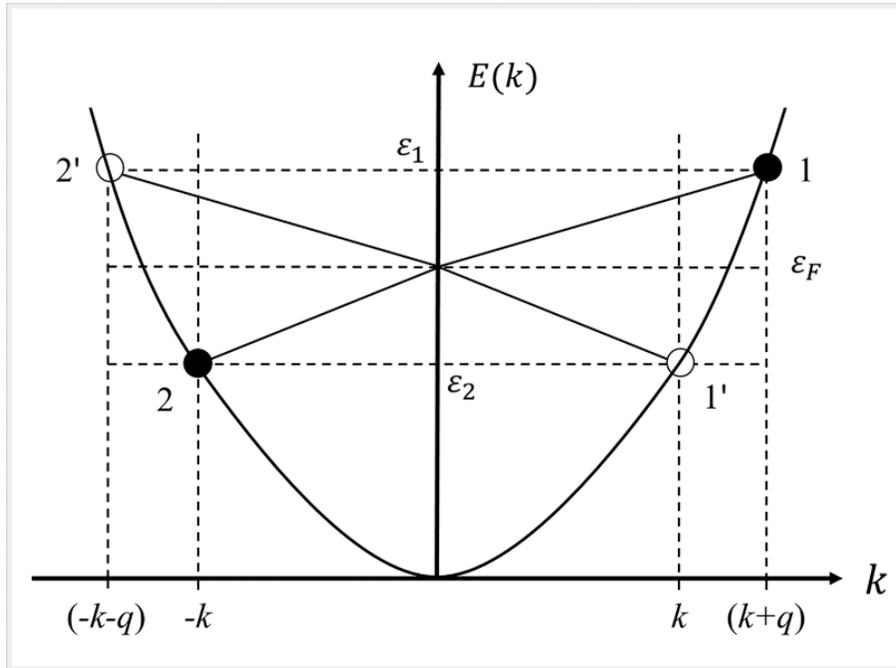

Figure 1. Conduction band diagram of a metal which reaches its critical temperature $T = T_C$

When $T = T_C$ is reached, the electrons are correlated in pairs that oscillate in Resonant Quantum States (RQS) by exchanging a phonon with energy $\hbar\omega_q$ and momentum $q$. At this moment, two interactions are connected: the coulombian repulsion and the electron-phonon-electron.

In Figure 1, when $T = T_C$, electron 1 (above the Fermi surface) with momentum $k+q$ and energy $\varepsilon(k+q)$ correlates with electron 2 (below the Fermi energy) with momentum $-k$ and energy $\varepsilon(-k)$. In addition, the energy difference between them is exactly the energy of the phonon.

When the coulombian repulsion (described by the potential $V^{e_1 e_2}$ and screened by the dielectric constant of Thomas-Fermi) is connected, electron 1 emits a phonon to electron 2 and now occupies the position of the hole 1' with momentum $k$ and energy $\varepsilon(k)$; meanwhile, electron 2 occupies the position of the hole 2' with momentum $-k-q$ and energy $\varepsilon(-k-q)$; this first process defines half oscillation. In the second half of the oscillation the opposite happens, the electrons in the 1' and 2' positions exchange a phonon and return to their original positions in 1 and 2. Note that, in the last exchange, the first electron absorbs a phonon (given by the potential $V^{e_1 f e_2}$) and the second electron emits a phonon (given by the potential $V^{e_1 \cdot f e_{2'}}$), which indicates that the potentials $V^{e_1 f e_2}$ and $V^{e_1 \cdot f e_{2'}}$ have opposite sign and are responsible for the complete static oscillation of the electrons, that is, in the absence of external field.

At the moment when $T = T_C$, the Hamiltonian of the system, Eq. (1), becomes

$$H_T = -\frac{\hbar^2 \nabla^2}{2m_e} + V_C + H_{int} \qquad (2)$$

here $H_{int}$ characterizes the internal state of the pair of electrons oscillating in the RQS, in the interval $0 \leq T \leq T_C$. Besides, $k$ is no longer a significant quantum number for the characterization of the internal state of the pair and the energies involved in it depend only on the temperature and the critical temperature. Then, in this temperature range, $H_{int}$ is the Hamiltonian with physical sense, because $H_T$ only participates when $T = T_C$.

Thus, we can say that

$$H_{int} = V^{e_1 e_2} \pm V^{efe} \qquad (3)$$

where the positive sign corresponds to $V^{efe} = V^{e_1 f e_2}$ for the first half oscillation and the negative sign corresponds to $V^{efe} = -V^{e_1 \cdot f e_{2'}}$ for the second half oscillation. And so, we can also say that $V^{e_1 f e_2} = -V^{e_1 \cdot f e_{2'}}$.

### a. Energy Band

At the moment in which $T = T_C$ (see Figure 1) the following relations are fulfilled

$$E(k+q) = E(-k-q) \text{ and } E(k) = E(-k)$$

as well

$$E(k+q) - E(k) = \hbar\omega_q \text{ and } E(-k-q) - E(-k) = \hbar\omega_q$$

representing the condition for oscillations. Then

$$\frac{\hbar^2(k+q)^2}{2m} + \frac{\hbar^2 k^2}{2m} = \hbar\omega_q$$

$$\frac{\hbar^2}{m}(k \cdot q) + \frac{\hbar^2 q^2}{2m} = \hbar\omega_q = \hbar cq \text{ with } k \cdot q = kq\cos\alpha(k,q)$$

however, for CEPO-RQS to exist when we have electrons that tend to equalize their momentum, but in the opposite direction, we must take into account that

$$\alpha(k,q) = \begin{cases} 0 & \text{if } q > 0 \\ \pi & \text{if } q < 0 \end{cases}$$

so

$$q\left(\frac{\hbar^2}{m}k + \frac{\hbar^2 q}{2m}\right) = \hbar cq \Rightarrow \frac{\hbar^2}{m}\left(k + \frac{q}{2}\right) = \hbar c$$

and finally we have

$$\left(k + \frac{q}{2}\right) = k_F \Rightarrow c = \frac{\hbar k_F}{m}$$

b. **Momentum**

The initial momentum of the correlated pair in the first half of the oscillations in the RQS is $k+q-k=q$, the momentum of the phonon emitted by the electron in position 1 and absorbed by the one in position 2. The momentum in the second half of the oscillations is $k-k-q=-q$, the momentum of the phonon emitted by the electron in the 2' position and absorbed by the one in the 1' position. The difference of the total momentum between the electrons in positions 1 and 2 and 1' and 2' is, respectively,

$$k+q+k=2k+q$$
$$-(k+q)-k=-2k-q$$

As we can see from these last two equations, the electrons oscillate in the RQS trying to equalize the momentum between them, individually, and change the sign of the momentum between pairs of states.

The total sum and difference between the momentum of the CEPO-RQS are the same in the first and second phases of the oscillation, but opposite, which justifies the oscillations of the pairs in these states since for each temperature in the interval $0 \leq T \leq T_C$ electrons have no other possibility of occupying other states, since these are the only ones allowed.

IV. **WAVEFUNCTIONS OF THE ELECTRONIC PAIRS CORRELATED IN THE RQS IN EVERY OSCILLATION**

At the moment when $T=T_C$ we can write the wave functions of the individual electrons 1 and 2 as $|k+q\rangle$ and $|-k\rangle$, and for the correlated pair in the RQS we

have $|k+q;-k\rangle$; the same happens with the individual holes 1' and 2', whose wave functions are $|-k-q\rangle$ and $|k\rangle$, and for the pair of holes in the RQS we have $|-k-q;k\rangle$ (see Figure 1).

## V. INTERNAL STATE CHARACTERIZATION OF THE ELECTRONIC PAIR IN THE RQS

For the calculation of the initial state of the oscillations in the RQS the pair formed by the electrons 1 and 2 with wave function $|k+q;-k\rangle$ we apply $H_{int}$

$$H_{int}|k+q;-k\rangle = \left(V^{e_1 e_2} + V^{e_1 f e_2}\right)|k+q;-k\rangle = \left(E^{e_1 e_2} + E^{e_1 f e_2}\right)|-k-q;k\rangle$$

where $E^{e_1 e_2}$ it is the energy corresponding to the coulombian repulsion operator between the electrons 1 and 2 and $E^{e_1 f e_2}$ is the energy that results from the exchange of the phonon emitted by the electron 1 and absorbed by the electron 2. Note that the wave function $|k+q;-k\rangle$ is not a self-function of the Hamiltonian $H_{int}$, so we apply $H_{int}$ again to obtain the initial wave function $|k+q;-k\rangle$

$$H_{int} H_{int}|k+q;-k\rangle = \left(V^{e_1 e_{2'}} + V^{e_{1'} f e_{2'}}\right)\left(V^{e_1 e_2} + V^{e_1 f e_2}\right)|k+q;-k\rangle$$
$$= \left(E^{e_1 e_2} + E^{e_1 f e_2}\right)\left(V^{e_1 e_{2'}} + V^{e_{1'} f e_{2'}}\right)|-k-q;k\rangle =$$
$$= \left(E^{e_1 e_2} + E^{e_1 f e_2}\right)\left(E^{e_1 e_{2'}} + E^{e_{1'} f e_{2'}}\right)|k+q;-k\rangle$$

In the last equation, we take into account that $V^{e_1 e_{2'}} = V^{e_1 e_2}$ and $V^{e_{1'} f e_{2'}} = -V^{e_1 f e_2}$, and if we consider $E^{e_1 e_2} = E^{e_1 e_{2'}}$ and $E^{e_1 f e_2} = -E^{e_{1'} f e_{2'}}$ we obtain

$$H_{int}^2 |k+q;-k\rangle = \left(V^{e_1 e_2} - V^{e_1 f e_2}\right)\left(V^{e_1 e_2} + V^{e_1 f e_2}\right)|k+q;-k\rangle$$
$$= \left(E^{e_1 e_2} - E^{e_1 f e_2}\right)\left(E^{e_1 e_2} + E^{e_1 f e_2}\right)|k+q;-k\rangle = \left(E^{e_1 e_2 \, 2} - E^{e_1 f e_2 \, 2}\right)|k+q;-k\rangle$$

The interaction Hamiltonian $H_{int}$ does not have self-states of the CEPO-RQS because it only characterizes half oscillation of the pair. It is $H_{int}^2$ the operator that characterizes the internal state of the CEPO-RQS for complete oscillation.

It is very important the interpretation of the square of the Hamiltonian as the square of the binding energy $(\Delta^2)$ between the CEPO-RQS in the range of $0 \leq T \leq T_c$. The participating energies $\Delta^2$, $(E^{e_1 e_2})^2$ and $(E^{e_1 f e_2})^2$ are functions that depend only on temperature and critical temperature between $0 \leq T \leq T_C$.

The dispersion relation of the CEPO-RQS in $0 \leq T \leq T_C$ is

$$\Delta^2 = \left(E^{e_1 e_2}\right)^2 - \left(E^{e_1 f e_2}\right)^2$$

Now we must find the explicit dependence of the energies $\Delta$, $E^{e_1 e_2}$ and $E^{e_1 f e_2}$ on the temperature and the critical temperature, for that we will analyze the instant in which $T = T_C$ it is reached. When we approach to $T_C$ from room temperature, the electrons are in individual states, while if we approach from $0K$ they are correlated.

When $T = T_C$, the electron-electron interaction is the only participant and the total energy of the electrons will depend on $T_C$. If we lower the temperature from $T_C$, then the electron-phonon-electron interaction is connected and the total energy will depend on $T$ for each $T_C$ and so

$$\Delta = \Delta(T; T_C), \ E^{e_1 e_2} = E^{e_2 e_1}(T_C) \ y \ E^{e_1 f e_2} = -E^{e_1 f e_{2'}}(T; T_C)$$

If we consider that $E^{e_1 e_2}$ has a linear dependence on temperature, then we can write

$$E^{e_1 e_2} = \rho k_B T_C \quad \text{and} \quad \left(E^{e_1 e_2}\right)^2 = \left(\rho k_B T_C\right)^2$$

where $\rho$ is the parameter that characterizes the qualitative nature of the material to achieve oscillations.

But the dependence on temperature is not linear for $E^{e_1 f e_2}$, it is more complicated because it has to do with a potential that involves at the same time the residual Coulomb repulsion and the interactions of both electrons with the phonon, and that is why we can propose a dependence as the following

$$E^{e_1 f e_2} = \rho k_B \sqrt{T_C T} \quad \text{and} \quad \left(E^{e_1 f e_2}\right)^2 = \left(\rho k_B\right)^2 T T_C$$

Now, the dispersion law that gives us the binding energy of the CEPO-RQS is

$$\Delta^2(T; T_C) = \left(\rho k_B T_C\right)^2 - \left(\rho k_B\right)^2 T T_C$$

or

$$\Delta(T; T_C) = \rho k_B T_C \left(1 - \frac{T}{T_C}\right)^{\frac{1}{2}}$$

This last equation coincides with that reported in Eq. (3.31) of the BCS article [1] and also coincides with the way Buckingham [10] suggests. The difference with our result is that in Eq. (3.31) of BCS [1] the parameter is considered constant and equal to 3.2 for all materials, but in our case it is different for each metal, it characterizes each substance to correlate the oscillation of electrons in the RQS as measured in the experiments.

If in the previous equation, we take $T \to 0K$, we obtain the parameter from the magnitudes that are measured experimentally

$$T \to 0K, \quad \rho = \frac{\Delta(0;T_C)}{k_B T_C}$$

which is constant, but different for each substance and we will call it correlation parameter. In Table 1 we present the correlation parameter for some metals and we can observe differences between them, we also present the corresponding values for $T_C$.

| ELEMENT | $T_C$ | $\Delta(0)/k_B T_C$ |
|---|---|---|
| Al | 1.2 | 3.4 |
| Cd | 0.56 | 3.2 |
| Hg (α) | 4.16 | 4.6 |
| In | 0.4 | 3.6 |
| Nb | 8.8 | 3.8 |
| Pb | 7.22 | 4.3 |
| Sn | 3.75 | 3.5 |
| Ta | 4.4 | 3.6 |
| Tl | 2.4 | 3.6 |
| V | 4.9 | 3.4 |
| Zn | 0.9 | 3.2 |

Table 1. $\Delta(0)$ is taken from tunneling experiments. $T_C$ and $\Delta(0)$ appear in R. Mersevey, B. B. (1969). Equilibrium Properties: Comparison of experimental results with predictions of the BCS Theory. At R. D. Parks, *Superconductivity vol I* (págs. 117-184). New York: Marcel Dekker Inc.

In the same way we can calculate the highest binding energy for a metal, which represents the energy accumulated when $T \to 0K$, obtaining

$$\Delta(0;T_C) = \rho k_B T_C$$

which means that as the temperature decreases from $T_C$ to zero, the energy accumulates as the binding energy of the pair.

Finally the equation that allows to calculate the binding energy of the pair as a function of the temperature and the critical temperature is

$$\Delta(T;T_C) = \Delta(0;T_C)\left[1 - \frac{T}{T_C}\right]^{\frac{1}{2}} \qquad (4)$$

## VI. DISCUSSION OF RESULTS

The correlation of two electrons is a matter of importance for physics because in the phenomenon of superconductivity the charge transport that has been measured experimentally is twice the charge of the electron, which suggests the existence of pairs of correlated electrons; but, on the other hand, we have the laws of Electromagnetic Theory, under which two electrons always repel themselves and are only forced to be drawn between them by an internal or external field.

Our model is based on obtaining the internal energy of the CEPO-RQS taking into account two interactions, the coulombian electron-electron and electron-phonon-electron, which is responsible for the pair oscillates. For this, we must reiterate that our pairs are conformed by three particles; Two fermions and one boson.

The study and consideration of the two interactions mentioned above results in an equation that provides the binding energy as a function of a parameter that identifies the nature of the superconductor, its temperature and its critical temperature.

The CEPO-RQS are pairs of electrons whose energy difference is equal to that of the phonon, and this is the necessary condition for them to be coupled in the oscillatory potential that governs the correlation; also, it is that condition that characterizes the pair in a resonant state, since both the phonon and the electron pair oscillate with the same frequency.

Equation (4) is a law of quadratic dispersion and is obtained after applying twice the Hamiltonian of interaction $H_{int}$ because we refer to an oscillatory potential where, if applied only once, it represents only half oscillation. This result coincides with the empirical law suggested by Buckingham [10] and corroborates the BCS in Eq. (3.31) of [1].

In the analysis of the parameter that we have called correlation parameter, we review its participation in plotting the binding energy for different materials and against the graph of the energy gap that appears in the literature observing, as shown in Figures 2 and 3, a different curve for each material that will not only depend on the variation in $T_C$, but on the variation of the parameter itself. In the literature, the correlation parameter appears as a constant whose value is 3.2 for all materials.

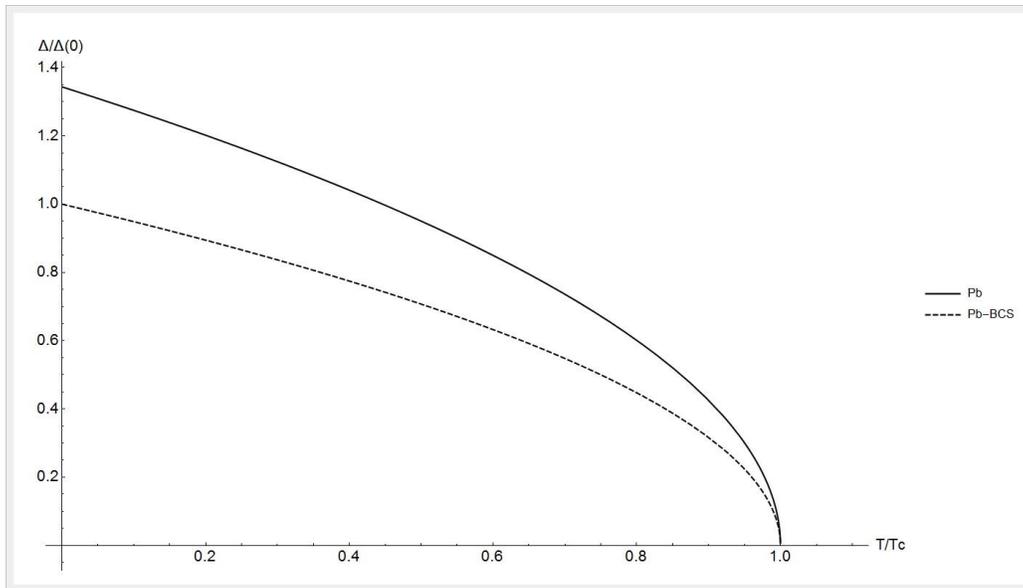

Figure 2. Variation for the $\Delta/\Delta(0)$ curves is shown after a $T/T_C$ normalization process, this is, comparing BCS Dispersion Law against the one obtained in this work, both calculated for Pb ($T_C = 7.22$, $\rho = 4.3$).

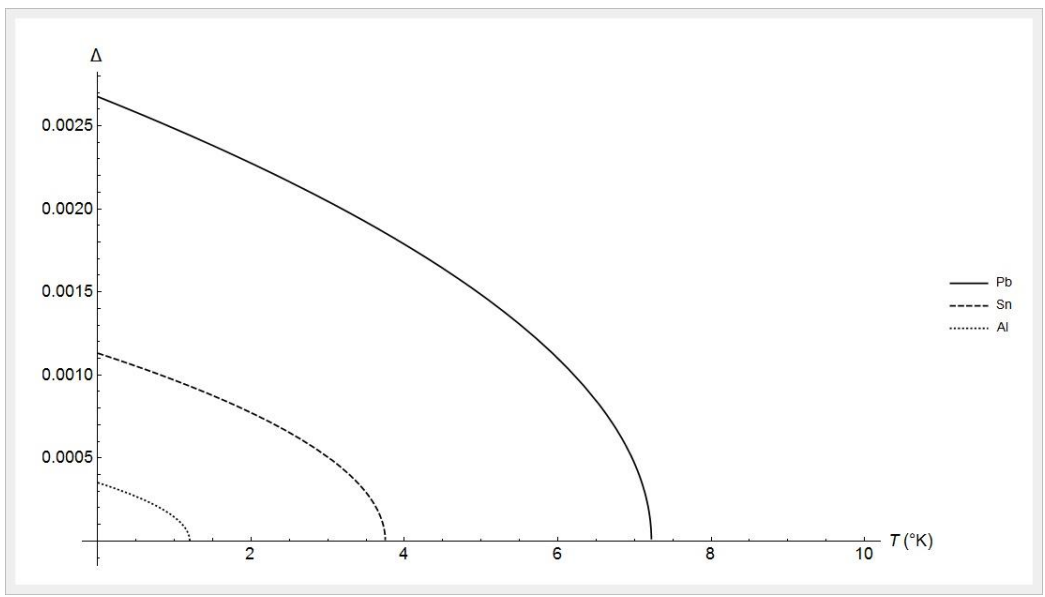

Figure 3. Variation for the $\Delta$ curves is shown while $\Delta$ is function of temperature, but also a function of the correlation parameter $\rho$.

## VII.  CONCLUSIONS

After an analysis of the results and in line with what has been reported in the reviewed literature, we consider that the model of the CEPO-RQS is of great importance for physics, and particularly for the theory of superconductivity.

In this theory, we describe pairs of electrons that differ from those presented by Cooper in his works which were considered in the BCS. The CEPO-RQS are pairs of electrons whose energy difference is equal to that of the phonon, and their momenta differ by a phonon momentum, although their signs are opposite, these are the necessary conditions for them to be coupled in the oscillatory potential that governs the correlation; also, it is that condition that characterizes the pair in a Resonant state, since both the phonon and the electron pair oscillate with the same frequency.

In the Cooper pairs of the BCS, it is conditioned that the difference of the energy of the electrons is greater than the one of the phonon and indicates a overscreening of the electron-electron interaction to form a pair of electrons which, contrary to repelling, attract.

On the other hand, we can conclude that the law of dispersion that is exposed in the theory BCS is not generic for all the metals, that is to say, the constant 3.2 that appears in the expression is not always the same, but is the parameter of correlation $\rho$ that varies for each material that acts together with the variation of the $T_c$. That is, our dispersion law includes that of the BCS when $\rho = 3.2$.

The Cooper Pair is a problem of two electrons overscreened because of the total dielectric function that includes electrons and phonons measured in the experimental observations of $2e$ transport and the relation $T_c \propto 1/M^\alpha$, proposed by Frölich.

CEPO-RQS are also fundamented in the same $2e$ charge transport and phonon interaction, but in this case is a system of three particules, two electrons and a phonon coupled, thus giving a new structure.

From the above, we can conclude that Cooper Pair is not the carrier of superconductivity as described in the BCS, but is a pair with the characteristics of the CEPO-RQS.